\documentclass[10pt,letterpaper]{article}
\usepackage[top=0.85in,left=1.5in,footskip=0.75in,marginparwidth=2in]{geometry}
\usepackage{amsmath}

\usepackage[utf8]{inputenc}

\usepackage{cite}

\usepackage{nameref,hyperref}

\usepackage{microtype}
\DisableLigatures[f]{encoding = *, family = * }

\usepackage{changepage}

\usepackage[aboveskip=1pt,labelfont=bf,labelsep=period,singlelinecheck=off]{caption}

\usepackage{lastpage,fancyhdr,graphicx}
\usepackage{epstopdf}
\fancyheadoffset[L]{2.25in}
\fancyfootoffset[L]{2.25in}

\usepackage{color}

\definecolor{Gray}{gray}{.25}

\usepackage{graphicx}

\begin{document}
\vspace*{0.35in}

\begin{flushleft}
{\LARGE
\textbf\newline{Spectral Peak Recovery in Parametrically Amplified THz-Repetition-Rate Bursts}
}
\newline

\textbf{Vinzenz Stummer\textsuperscript{1,*},
Tobias Flöry\textsuperscript{1},
Matthias Schneller\textsuperscript{1},
Edgar Kaksis\textsuperscript{1},
Markus Zeiler\textsuperscript{1},
Audrius Pugžlys\textsuperscript{1,2},
Andrius Baltuška\textsuperscript{1,2}
}
\\
\bigskip
[1] Photonics Institute, TU Wien, Gusshausstrasse 27/387, 1040 Vienna, Austria
\newline
[2] Center for Physical Sciences \& Technology, Savanoriu Ave. 231 LT-02300 Vilnius, Lithuania
\\
\bigskip
* vinzenz.stummer@tuwien.ac.at

\end{flushleft}

\section*{Abstract}
Multi-photon resonant spectroscopies require tunable narrowband excitation to deliver spectral selectivity and, simultaneously, high temporal intensity to drive a nonlinear-optical process. These contradictory requirements are achievable with bursts of ultrashort pulses, which provides both high intensity and tunable narrowband peaks in the frequency domain arising from spectral interference. However, femtosecond pulse bursts need special attention during their amplification [Optica 7, 1758 (2020)], which requires spectral peak suppression to increase the energy safely extractable from a chirped-pulse amplifier (CPA). Here, we present a method combining safe laser CPA, relying on spectral scrambling, with a parametric frequency converter that automatically restores the desired spectral peak structure and delivers narrow linewidths for bursts of ultrashort pulses at microjoule energies. The shown results pave the way to new high-energy ultrafast laser sources with controllable spectral selectivity.


\section{Introduction}
\begin{figure}
\centering
{\includegraphics[width=\linewidth]{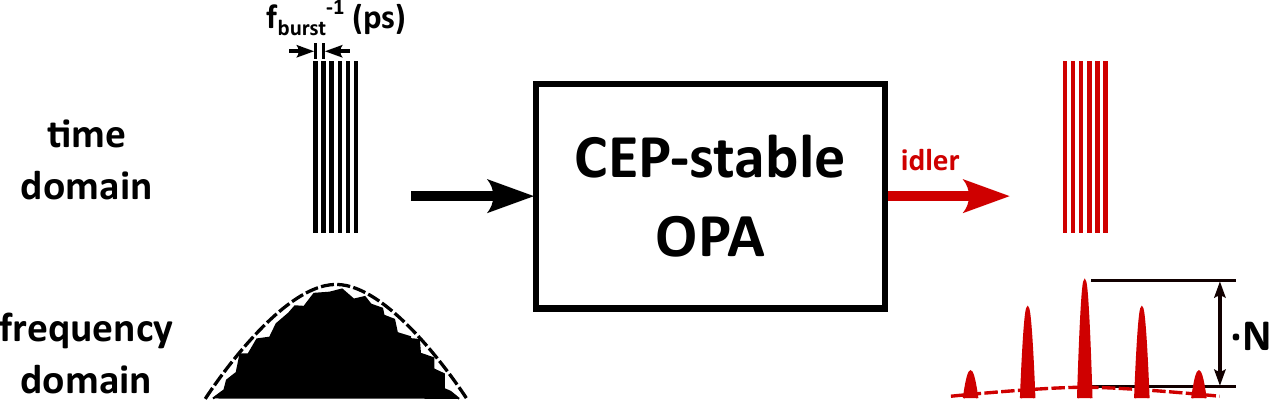}}
\caption{Concept of utilizing a passive CEP-stabilizing OPA to recover the spectral peak structure of a burst of ps-spaced ultrashort pulses amplified via PSCPA by the generation of a $\mu$J phase-descrambled idler burst. Bursts are outlined in the time- and frequency domain. In the frequency domain, the single pulse spectrum is depicted as a dashed line and the burst spectrum as a solid line with a filled area underneath. On the left side, the phase-scrambled spectrum shows a resemblance to the single pulse spectrum because the spectral interference is suppressed. On the right side, the phase-descrambled spectrum shows an $N$-proportional increase in spectral peak intensity compared to the spectrum of a single pulse with the same total energy.}
\label{fig:intro}
\end{figure}

In the past decades, finite trains of ultrashort pulses, also known as femtosecond (fs) pulse bursts, have become an increasingly valuable pulse format. Given with an intraburst repetition rate of some gigahertz (GHz) or lower, they have found their use in a large number of applications, such as materials processing \cite{kerse_ablation-cooled_2016}, pulsed laser deposition \cite{murakami_burst-mode_2009}, laser-induced breakdown spectroscopy \cite{skruibis_multiple-pulse_2019}, and seeding of free-electron lasers \cite{sudar_burst_2020}. Furthermore, there is a growing number of applications, which would highly benefit from a burst of ultrashort pulses where the intraburst repetition rate is increased to its limit, such that the pulse spacing becomes comparable to the duration of individual pulses. Sub-picosecond (ps) pulses correspond to terahertz (THz) intraburst repetition rates or, equivalently, ps spacings in time. This demand is given by the fact that many molecular dynamics due to their (ro-)vibrational time scales lie within the ps range. For example, the coherent control of rotational wavepackets can be well improved with already a handful of ps-spaced pulses \cite{schroeder_molecular_2020, lee_two-pulse_2004, pabst_alignment_2010}. In the frequency domain, a burst spectrum of $N$ equal pulses shows a strong spectral peak structure with a period given by the intraburst repetition rate. This structure originates from spectral interference of the $N$ burst pulses with an $N$-proportional increase in burst spectral intensity at a given energy, in contrast to if the entire burst energy were put into a single pulse ($N$ = 1). This allows for a frequency-selective format of ultrashort pulses, which is expected to strongly increase the efficiency of several nonlinear spectroscopic applications, such as Stimulated Raman Scattering (SRS) \cite{prince_stimulated_2017}, Resonantly-Enhanced Multi-Photon Ionization (REMPI) \cite{zhang_coherent_2007} and Infrared Resonant Multi-Photon Dissociation (IRMPD) \cite{ligare_resonant_2015, wellers_resonant_2011, nieckarz_infrared_2013}.\\
To be a useful tool, ultrashort pulses need to be amplified to at least sub-millijoules for these applications. While for an intraburst repetition rate of some GHz or lower, amplification is a straightforward task, the application of chirped-pulse amplification (CPA) \cite{strickland_compression_1985} to THz-repetition-rate bursts leads to several difficulties, such as a strong reduction of the reachable energy and potential amplifier damage. This is because of the translation of the burst spectrum into the time domain when amplifying a handful of ps-spaced, strongly chirped pulses in an amplifier when the stretched pulse duration strongly exceeds the duration of the burst. Overcoming the mentioned technical limitations, we were able to demonstrate THz-repetition-rate burst generation in the near-infrared (NIR) at multiple millijoules of burst energy \cite{stummer_programmable_2020}. This was enabled by the application of phase scrambling, where the relative carrier-envelope phases (CEP) of the burst pulses (phase scrambling) were pseudo-randomly modulated, which leads to a smearing of the burst-typical spectral peak structure. This mimics the spectral properties of a single pulse and thus avoids the breakdown of the CPA approach, and extends it by combining it with pulse phase modulation techniques. We refer to it as phase-scrambled CPA (PSCPA) from this point on.\\
However, while the suppression of spectral peaks allowed for mJ burst amplification, it removes the capability to be spectrally selective when applying amplified bursts. In this work, we demonstrate the recovery of the burst spectral peak structure after PSCPA by parametric amplification. As outlined in Fig. \ref{fig:intro}, we drive an Optical-Parametric Amplifier (OPA), that is capable of passive CEP stabilization, with a burst that is amplified to sub-mJ energies by application of PSCPA. For individual pulses, the concept of passively stabilizing the CEP with OPAs is already widely known \cite{baltuska_controlling_2002}. It is based on the canceling of signal and pump pulse CEPs when being imprinted on the idler signal. In the present work, the second harmonic of the driving burst serves as a pump and generates a white-light signal seed. This leads to a $\mu$J idler signal with phase-descrambled burst pulses and a recovery of the spectral peak structure.

\section{Regime of Phase-Scrambled CPA (PSCPA)}
\label{sec:regime}
\begin{figure}[htbp]
\centering
{\includegraphics[width=\linewidth]{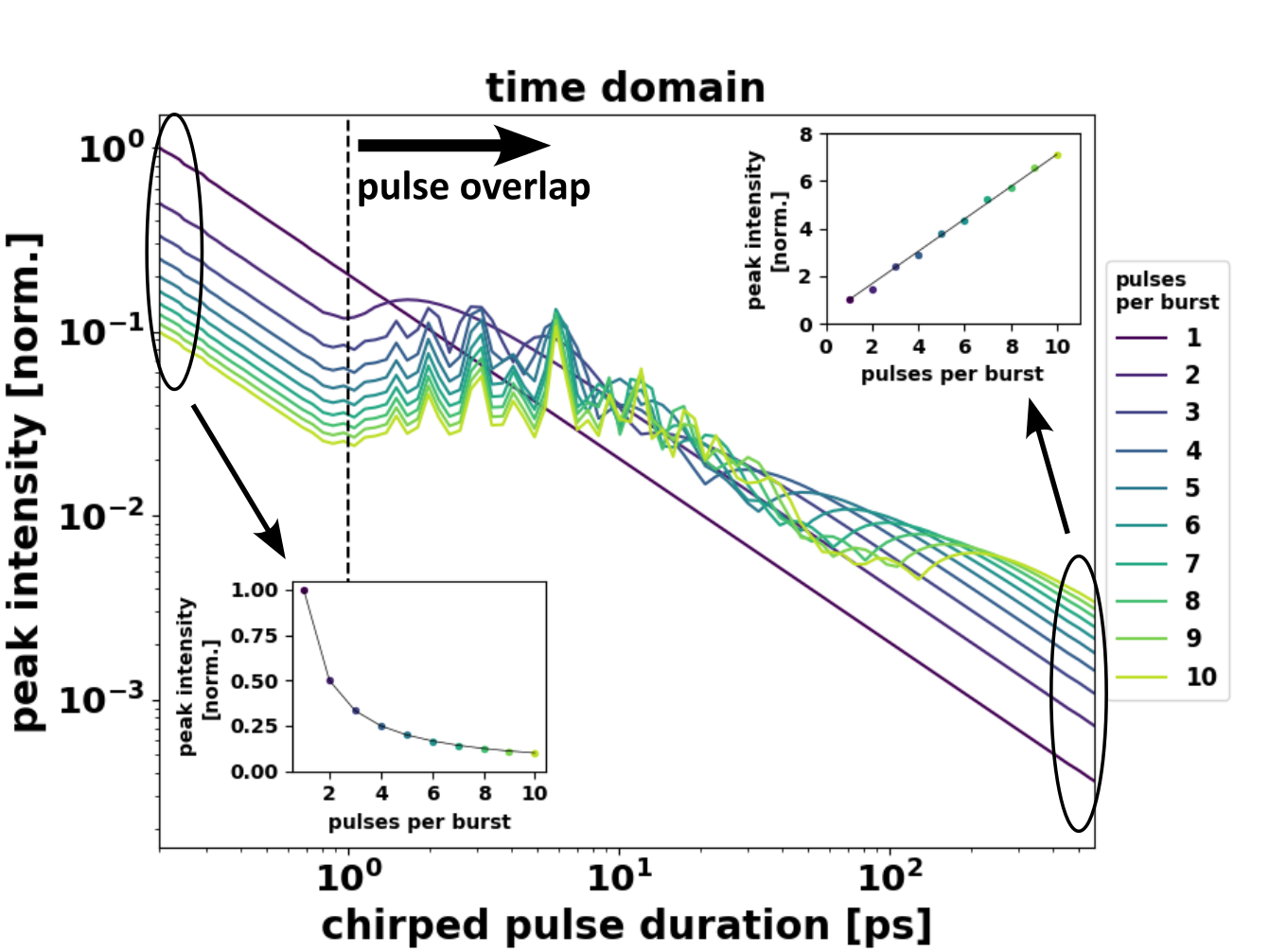}}
\caption{Maximum temporal peak intensity for a burst of chirped pulses with a pulse spacing of 1 ps, depending on the chirped pulse duration. The plot is shown for various pulse numbers while keeping the total burst energy constant, with temporal peak intensity being normalized to the compressed single pulse case. Lower left inlay: Maximum temporal peak intensity (normalized to the corresponding single-pulse case) vs. pulse number in the case of fully compressed 250 fs pulses. Upper right inlay: Maximum temporal peak intensity (normalized to the corresponding single-pulse case) vs. pulses number in the case of pulses stretched to 600 ps.}
\label{fig:map}
\end{figure}

Before we describe the experiment and communicate its results, we would like to address the regime in which PSCPA becomes relevant. In the frequency domain, a burst of $N$ pulses with pulse spacing $\Delta t$ can be described by the product of its single pulse complex envelope $\tilde{E}_P(\omega)$ and a factor $\tilde{f}(\omega;\Delta t,N,\phi_{slip})$ describing the spectral interference \cite{oppenheim_signals_1996}:

\begin{align}
\label{eq:1}
\tilde{E}_B(\omega) &= \tilde{E}_P(\omega) \cdot \tilde{f}(\omega;\Delta t,N,\phi_{slip})\nonumber\\ &= \tilde{E}_P(\omega) \cdot \sum_{n=0}^{N-1}\exp{\left(-in(\Delta t \omega - \phi_{slip})\right)}\nonumber\\ &= \tilde{E}_P(\omega) \cdot \frac{
										\sin\left(\frac{N(\Delta t \omega-\phi_{slip})}{2}\right)}
										{\sin\left(\frac{\Delta t \omega-\phi_{slip}}{2}\right)}
								  e^{-j\frac{N-1}{2}(\Delta t \omega-\phi_{slip})},
\end{align}

with $\phi_{slip}$ being the pulse-to-pulse phase slip in the burst. For identical burst pulses, the squared absolute maximum of the interference factor $|\tilde{f}(\omega;\Delta t,N,\phi_{slip})|^2$ is equal to $N^2$. Equivalently speaking, the burst spectral peak intensity is $N$-times higher for a burst with $N$ pulses at a given total burst energy $E_B$ than a single pulse ($N$=1) with the same energy $E_B$. This enables one to increase the spectral peak intensity with a factor of $N$ at a given energy $E_B$, while also becoming spectrally selective for sample coherence times that are larger than the burst duration.\\
For sufficiently strongly chirped pulse durations compared to the temporal duration of the compressed burst $(N-1)\Delta t$, the burst spectrum is well present in the time domain. Assuming linearly chirped pulses

\begin{equation}
    \label{eq:chirp}
    E_P(t) = E_0 \cdot \exp{
    \left(2\ln{(2)}(1+iC_t)(t/\tau_{ch})^2\right)
    },
\end{equation}

with amplitude $E_0$, chirp parameter $C_t$, and chirped pulse duration (FWHM) $\tau_{ch}$, we calculated numerically the burst temporal intensity using Fourier transform methods together with Eqs. (\ref{eq:1}) and (\ref{eq:chirp}). From the compressed pulse case up to strongly chirped pulses with hundreds of ps duration, we calculated the temporal peak intensity for up to $N=10$ pulses with 1 ps spacing, while keeping the total energy constant. The results can be seen in Fig. \ref{fig:map}. The sub-ps duration range determines the regime where divided-pulse amplification (DPA) is applied \cite{zhou_divided-pulse_2007,klenke_530_2013,kienel_analysis_2013} and is indicated by a 1/$N$-decrease of the temporal peak intensity (see Fig. \ref{fig:map}, lower left inlay) by an equal distribution of the total energy among the burst pulses. In the other case, for strongly stretched pulses with chirped pulse durations of several hundreds of picoseconds, the behavior is the opposite: by increasing pulse number, we see an $N$-times increase in the temporal peak intensity compared to the single pulse case (see Fig. \ref{fig:map}, upper right inlay), which identifies the PSCPA regime. In this region, the burst spectral peaks are well present in the time domain and thus decrease amplifier reachable energy and stability.

\newpage

\section{Experimental Setup}
\subsection{Overview}
\label{sec:experimental}

\begin{figure*} [htbp]
{\includegraphics[width=\linewidth]{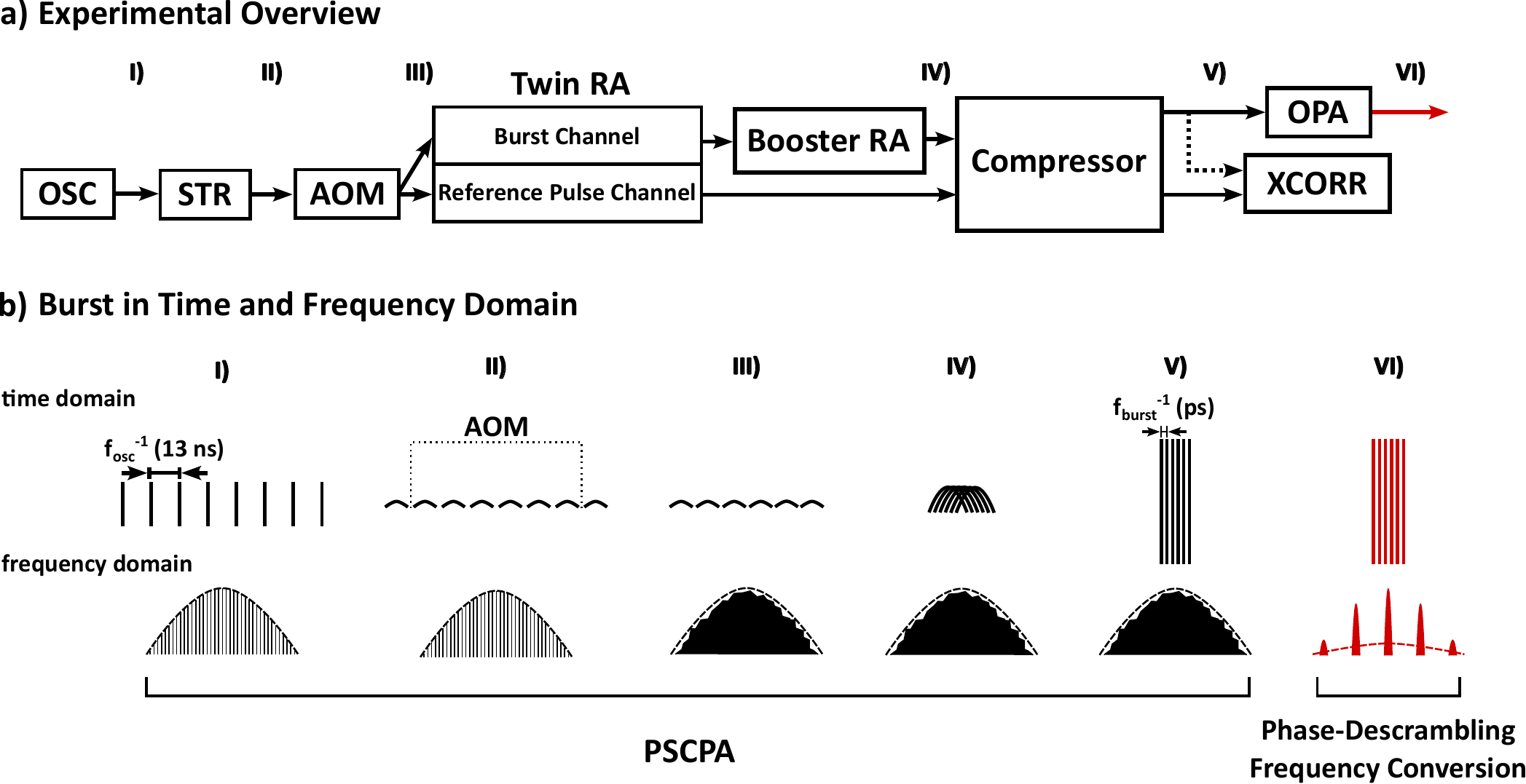}}
\caption{a) Experimental Overview. OSC: Mode-locked MHz Oscillator. STR: Stretcher. AOM: Acousto-Optic Modulator. RA: Regenerative Amplifier. OPA: Optical Parametric Amplifier. XCORR: Cross-Correlation Stage. PSCPA: Phase-Scrambled Chirped Pulse Amplification. b) Burst Hybrid Amplification in the time and frequency domain. Single-pulse spectra are represented as dashed lines, and multi-pulse spectra as solid lines with a filled area underneath.}
\label{fig:experimental}
\end{figure*}

The full experimental overview of this work can be seen in Fig. \ref{fig:experimental}(a), together with a temporal and spectral visualization of the burst labeled with Roman digits in between each of the stages in Fig. \ref{fig:experimental}(b). In general, the amplification chain consists of a PSCPA part (Fig. \ref{fig:experimental}(b) I-V) followed by a phase-descrambling spectral peak recovery part represented by the OPA (Fig. \ref{fig:experimental}(b) V-VI). A MHz-repetition-rate mode-locked oscillator (1030 nm Yb:KGW, 76 MHz repetition rate, 250 fs pulse duration) generates nanojoule pulses, which correspond to narrowband, MHz-spaced spectral comb teeth below the single-pulse envelope (Fig. \ref{fig:experimental}(b) I). The oscillator pulses are stretched to around 300 ps by a double-pass grating stretcher. The stretcher mainly imprints a spectral phase to stretch the pulses (Fig. \ref{fig:experimental}(b) II). Because of gain narrowing during amplification, precompensation by shaping the pulse spectrum in the Fourier plane of the stretcher is also applied but not depicted in the Figure due to clarity. An acoustic-optical modulator (AOM) diffracts the burst seed pulses and provides individual amplitude and phase modulation of AOM-diffracted pulses. Here, we apply phase scrambling by suitable pseudo-random pulse phase modulation \cite{stummer_programmable_2020}, such that constructive spectral interference is suppressed and the multi-pulse spectrum resembles the single pulse spectrum (\ref{fig:experimental}(b) III). Amplification takes place at a repetition rate of 1 kHz. For being able to determine individual burst pulse energies after amplification, we amplify in parallel to the burst a single pulse as a reference for direct temporal burst characterization. This enables us to equalize burst pulse energies by gain precompensation carried out by amplitude modulation in the AOM. Strong variations in burst pulse energies would otherwise deteriorate the parametric conversion performance of the burst in the OPA since all relative intensity variations are increased further by nonlinear processes. For this we built up a CW-pumped twin regenerative amplifier (twin RA, Yb:CaF$_2$) with two cavities: in one we accumulate and pre-amplify the AOM-diffracted pulses to a burst of pulses with ps spacing and 100 $\mu$J of burst energy, in the other we amplify the single reference pulse to 50 $\mu$J. The prior requires that the round-trip time of the burst cavity is comparable to the oscillator round-trip time, such that their absolute difference gives the intraburst pulse spacing (Vernier effect). By application of an intermediate voltage to the RA Pockels Cell, the transparency of the RA cavity can be set such that burst seed pulses can couple into the cavity without being fully ejected during burst buildup. For further details on PSCPA and the burst formation and amplification process, we would like to refer again to our previous publication, where these aspects are covered in more detail \cite{stummer_programmable_2020}. Since both cavities are seeded by the same oscillator, the burst and the reference are thus synchronized to each other. The burst is amplified further up to 200 $\mu$J for each burst pulse (Fig. \ref{fig:experimental}(b) IV) in a home-built cryogenically cooled Yb:CaF$_2$ booster amplifier. The burst pulses and the reference pulse are then recompressed (Fig. \ref{fig:experimental}(b) V) to a pulse duration of 250 fs in a compressor where both channels are double passing through a single 130x20 mm$^2$ transmission grating (Gitterwerk) while being spatially separated from each other. Losses in the compressor have been determined to be around 30$\%$, resulting in a compressed burst pulse energy of 140 $\mu$J and a compressed reference pulse energy of 35 $\mu$J. We generate bursts consisting of 6 pulses, which results in a compressed burst energy of 840 $\mu$J. The compressed pulse burst is then sent into a passive CEP-stabilizing OPA for spectral peak recovery. The idler output of the OPA gives a burst of ultrashort pulses at $\mu J$ energies with a clean spectral modulation coming from the spectral interference of phase-descrambled burst pulses (Fig. \ref{fig:experimental}(b) VI). A small part (few $\mu$J) of the PSCPA burst is sent together with the reference pulse to a cross-correlation stage for the determination of individual burst pulse energies. For the cross-correlation measurement, the temporal envelope of the whole burst is measured by scanning the time delay between the burst and the reference pulse with a mechanical delay line. As a signal we detect the sum frequency signal generated in a type I $\beta$-barium borate (BBO) nonlinear crystal during the overlap of the burst with the reference pulse under a few-degree angle.

\subsection{Phase-Descrambling Frequency Conversion in an OPA}
\begin{figure}[htbp]
\centering
{\includegraphics[width=\linewidth]{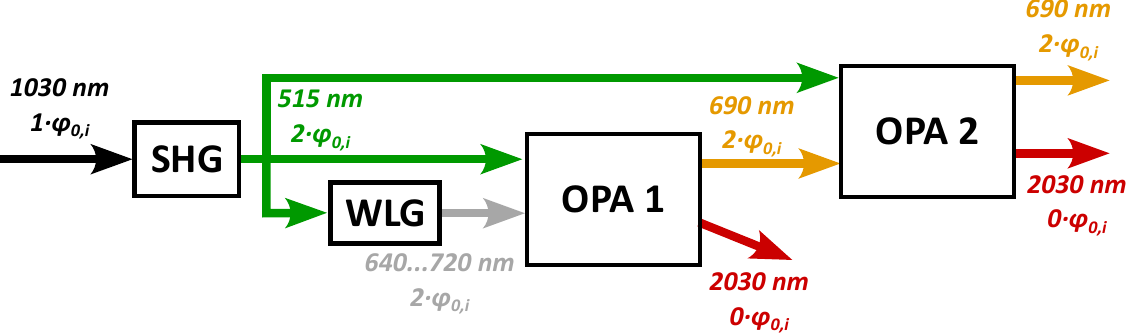}}
\caption{2-stage OPA scheme for phase-descrambling and frequency conversion. SHG: Second-Harmonic Generation. WLG: White-Light Generation. OPA: Optical-Parametric Amplifier.}
\label{fig:OPA}
\end{figure}

The OPA represents a phase-descrambling frequency conversion stage, in which the phase difference between pump and signal is imprinted on the idler. The burst spectral peaks and the scrambling of phases in PSCPA regard only the relative CEP $\phi_{0,i}$ that is given by the CEP of burst pulse $i$ relative to the first burst pulse. In contrast, the OPA can passively stabilize the total CEP of burst pulses and therefore also the relative CEP $\phi_{0,i}$ of burst pulses with respect to each other. When the CEPs of each pump and signal pulse pair are equal, the idler pulses are all expected to have the same CEP. This is the burst equivalent of a well-known process for individual pulses \cite{baltuska_controlling_2002}, whose single-pulse description is completely applicable to this case since the individual burst pulses are well separated from each other in time, given a pulse duration of about 250 fs and a pulse spacing of more than 1 ps. This way, passive CEP stabilization demodulates individual burst pulses in their phase and thus allows for a recovery of the burst spectral peak structure that is lost during PSCPA. Further, this concept of OPA phase descrambling enables conversion of the carrier frequency from the NIR, where highly efficient amplification is possible, to any OPA-reachable desired frequency.\\
Fig. \ref{fig:OPA} shows a conceptual scheme of the passive CEP-stabilizing OPA used for the experiments in this work. As mentioned in the section before, the OPA is driven by an NIR burst consisting of 6 pulses, each amplified to 140 $\mu$J. We use the second harmonic of the NIR input burst both as the pump for the OPA stages and generation of the WLG seed burst. The second-harmonic conversion efficiency is about 35$\%$ resulting in a 50 $\mu$J second-harmonic burst. An increase of the individual burst pulse energy is not used because of the onset of parametric superfluorescence in the OPA stages at higher pump energies. Due to the use of the second harmonic, there are twice the individual pulse relative phases $\phi_{0,j}$ each imprinted on the pump pulses and the WLG seed pulses. We tune the first-stage OPA crystal (BBO Type I) to an angle to achieve phase matching at around 690 nm. While the signal of the first stage acquires twice the original individual pulse relative phases, the idler has a constant CEP for all burst pulses ($\phi_{0,j}$=0). However, to get a sufficient signal-to-noise ratio (SNR), we operate the first stage non-collinearly, separate spatially the signal from the idler burst by a few-degree angle and further amplify the signal in a collinear second stage (BBO Type I). We then measure the seed pulse spectrum and the spectra of the second-stage signal at around 690 nm and of the idler at 2030 nm, with idler energies being a few $\mu$J.

\section{Results}

For all measurements, we performed cross-correlations and acquired spectra. At first, we measured with a single pulse for initial OPA alignment and signal optimization. We then measured with a burst of 6 pulses and a burst pulse spacing of 1.63 ps. Finally, we adjusted the burst cavity round-trip time in the twin RA burst channel by using a micrometer screw that finely tunes the position of the cavity end mirror. By this, we got a burst pulse spacing of 1.99 ps and carried out another measurement. 

\subsection{Cross correlations and spectra}
\begin{figure*}
\centering
{
\includegraphics[width=0.8\linewidth]{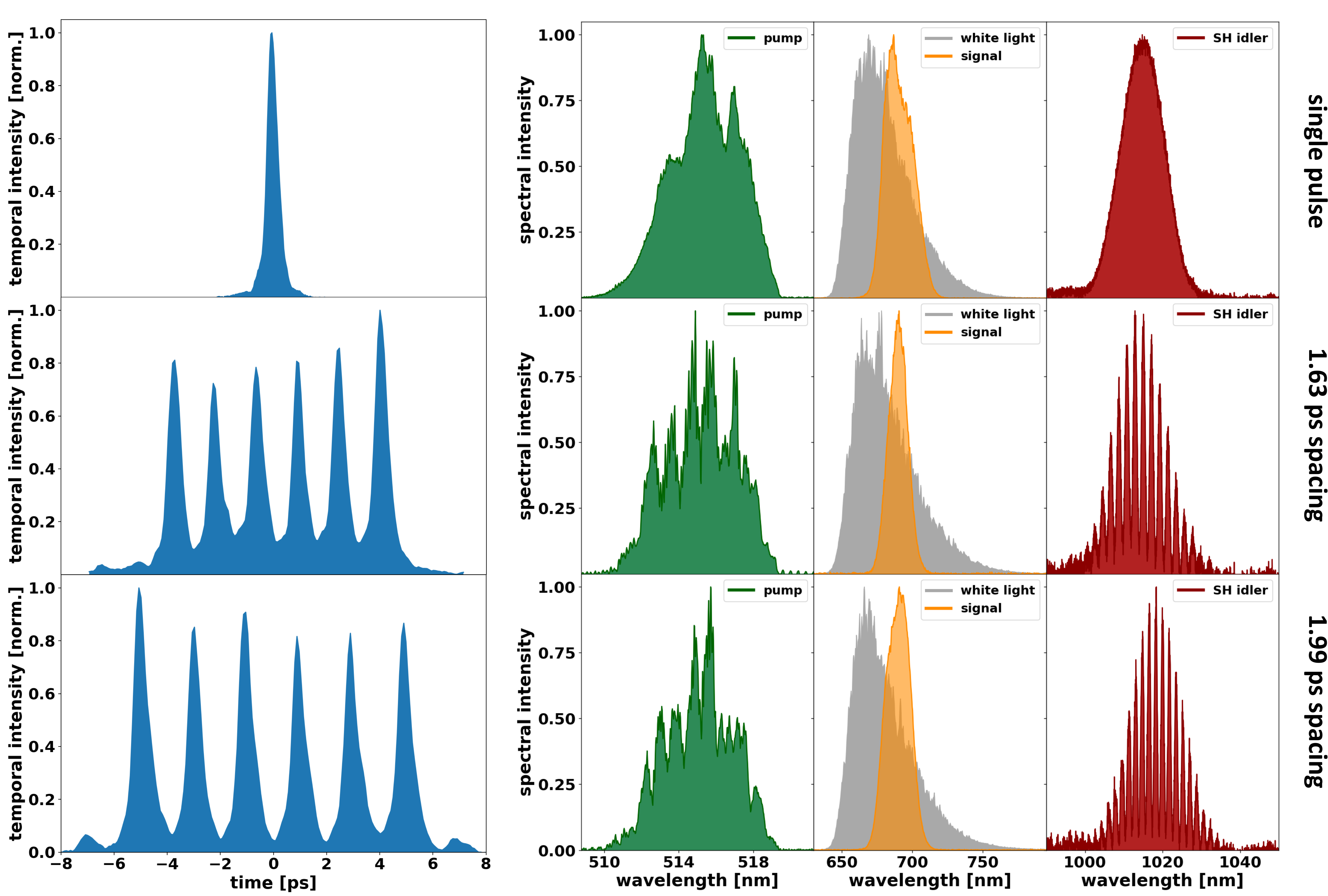}
}
\caption{Cross Correlations (left, blue) and spectra (other) acquired for a single pulse (top) and a burst of 6 pulses with a temporal spacing of 1.63 ps (middle) and 1.99 ps (bottom). The white-light-generated seed spectrum (light orange) and the second-stage signal spectrum (orange) can be seen. The second-harmonic of the idler signal generated in the second stage (red) is also visible, showing a clean periodic modulation, originating from the phase-descrambling of the burst pulses.}
\label{fig:spectra}
\end{figure*}

In Fig. \ref{fig:spectra}, the results consisting of cross-correlations and spectra of pump, WLG seed, 2-stage signal, and idler are shown. The pulse duration of 250 fs is confirmed by the cross-correlation measurement, as is the pulse spacing of 1.63 ps and 1.99 ps in the burst cases. For the idler, we measured its second harmonic in a single-shot manner with an integration time of 1 ms (Avantes Avaspec-ULS4096CL-EVO) synchronously to the burst amplification repetition rate of 1 kHz. In all experiments, we see that the pump spectrum coming from the phase-scrambled burst has a stochastic structure without any pronounced burst modulation. It is confirmed by the measurement that, as outlined in the section before, the phase scrambling is transferred to both the WLG seed and the signal (Ocean Optics HR4000/HR4000CG-UV-NIR), which thus lacks any sign of periodic modulation with a period according to the pulse spacing. In agreement with the theoretical predictions, the burst idler spectra consist of a clear periodic modulation with high modulation depth resulting from spectral interference of phase-descrambled burst pulses according to their temporal spacing. Burst idler spectral periods can be determined to be 2.1 nm and 1.7 nm, which at 1015 nm (SH of idler) corresponds well to the given pulse spacings of 1.63 ps and 1.99 ps, respectively. As expected for a single pulse, spectral interference modulations are completely absent in its idler spectrum. 

\subsection{Comparison with the theoretical limit}

\begin{figure}[htbp]
\centering
\includegraphics[width=0.75\linewidth]{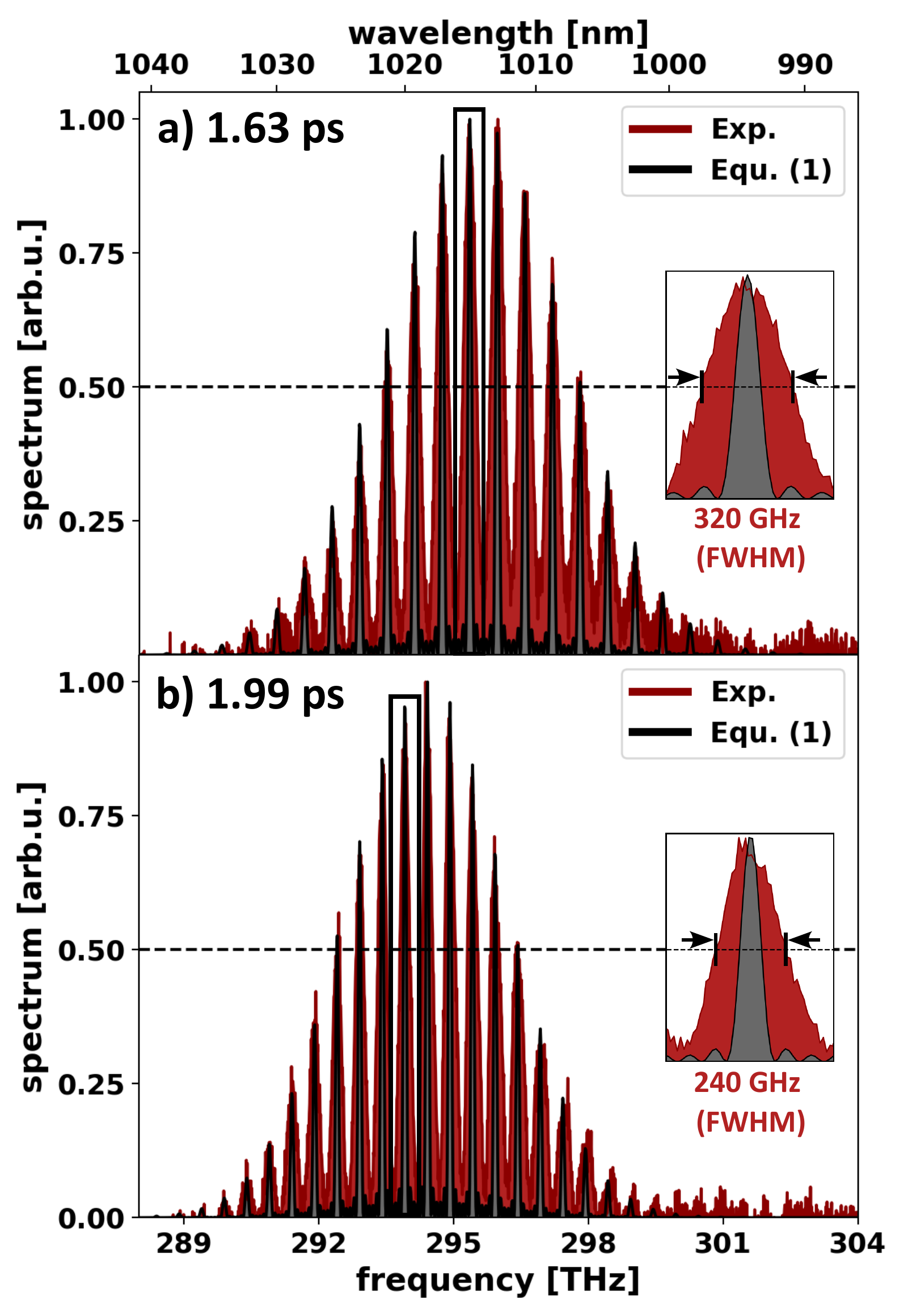}
\caption{Comparison of the measured SH idler spectrum (red) with the calculated fit (black) according to Equ. (1) for a pulse spacing of a) 1.63 ps and b) 1.99 ps.}
\label{fig:comp}
\end{figure}

To further confirm our assumption of the origin of the idler spectral modulation, we compare the measured SH idler spectra with an analytical calculation. For a burst of $N$ pulses, we can calculate the burst spectral intensity from Eq. (\ref{eq:1}). From this, we performed a fit for each of the measurements and plotted the results together with the measured spectra in Fig. \ref{fig:comp}.\\
Overall, we can see that the comparison looks very promising, with a broadening of experimentally measured spectral peaks compared to the theoretically calculated ones: For the short/long pulse spacing case, we determined the peak width (FWHM) to be 320/240 GHz in the measurement compared to 90/80 GHz according to the calculation. This broadening can be explained by the fact, that the idler is angularly chirped which is also the reason why after optimization via alignment of the beam into the spectrometer the central frequency can differ by a few THz. This effect is noticeable by a change in the measured central frequency from 295.5 THz at 1.63 ps spacing to 294.4 THz at 1.99 ps spacing. Since an angular chirp also leads to time smearing \cite{martinez_pulse_1986}, this smearing effect also explains the broadening of the spectral peaks that are formed by the very short pulse spacing \cite{cirmi_novel_2020}.

\newpage

\section{Conclusion and Outlook}
We have successfully demonstrated the recovery of the burst spectral peak structure after amplification of a THz-repetition-rate burst, consisting of 6 pulses, to sub-mJ energies via PSCPA. We did so by driving a passive CEP-stabilizing OPA with the PSCPA burst, where by suitable OPA design the initial phase-modulation could be undone in the burst idler. This work demonstrates a burst equivalent of the already known passive CEP-stabilizing effect of single idler pulses \cite{baltuska_controlling_2002}, confirming that closely spaced pulses still act as individual pulses when applying nonlinear optical conversion techniques, such as SFG, DFG, and WLG. This aspect confidently shows the usefulness of THz-repetition-rate bursts for the application of various nonlinear optical applications that rely on these techniques. The recovered spectral peaks can be controlled by changing sensitively the burst pulse spacing for its use in nonlinear spectroscopic applications that require wavelength scanning. Due to the high SNR of the spectral results, single-shot measurements can also be already carried out with this technique, allowing for fast data acquisition. Regarding the observed burst spectral peak broadening, we note that there is a growing number of compensation techniques for angularly chirped pulses available \cite{cirmi_novel_2020,hu_angular_2022}, that could be applied in the future. \\
Overall, this progress in high-energy THz-repetition-rate burst technology enables several new developments in the coherent control of (ro-)vibrational wavepackets and in nonlinear spectroscopy, with several useful applications, such as the alignment of molecules, non-dispersive absorption measurements, SRS and REMPI.

\section*{Funding}
Österreichische Forschungsförderungsgesellschaft (I 4566); Hochschuljubiläumsstiftung der Stadt Wien (H-260716/2020).

\section*{Disclosures}
The authors declare no conflicts of interest.

\bibliographystyle{unsrt} 
\bibliography{article}

\end{document}